\documentclass[useAMS,usenatbib]{mn2e}

\RequirePackage[dvips]{graphicx}

\def\aap{\emph{A.\& A.}}

\def\apj{\emph{ApJ.}}
\def\apjs{\emph{ApJ. Supp.}}
\def\apjl{\emph{ApJ. Lett.}}

\def\mnras{\emph{MNRAS}}

\title[Frequency-Luminosity Relation]{A New Frequency-Luminosity Relation for Long GRBs?}

\author[Ukwatta et al.]{T. N. Ukwatta$^{1,2}$\thanks{E-mail:
tilan.ukwatta@gmail.com (AVR)}, K. S. Dhuga$^{1}$, D. C.
Morris$^{1,2}$, G. MacLachlan$^{1}$, W. C. Parke$^{1}$, \newauthor
L. C. Maximon$^{1}$, A. Eskandarian$^{1}$, N. Gehrels$^{2}$, J. P. Norris$^{3}$, and A. Shenoy$^{1}$\\
$^{1}$Department of Physics, The George Washington University, Washington, D.C. 20052, USA.\\
$^{2}$NASA Goddard Space Flight Center, Greenbelt, MD 20771, USA.\\
$^{3}$University of Denver, Department of Physics and
Astronomy, 2112 East Wesley Ave. Room 211, Denver CO 80208, USA.\\
}

\begin{document}



\maketitle

\label{firstpage}

\begin{abstract}
We have studied power density spectra (PDS) of 206 long Gamma-Ray
Bursts (GRBs). We fitted the PDS with a simple power-law and
extracted the exponent of the power-law ($\alpha$) and the
noise-crossing threshold frequency ($f_{\rm th}$). We find that
the distribution of the extracted $\alpha$ peaks around $-1.4$ and
that of $f_{\rm th}$ around 1 Hz. In addition, based on a sub-set
of 58 bursts with known redshifts, we show that the
redshift-corrected threshold frequency is positively correlated
with the isotropic peak luminosity. The correlation coefficient is
$0.57 \pm 0.03$.
\end{abstract}


\section{Introduction}\label{Introduction}

Gamma-ray Bursts (GRBs) show very complicated time profiles and,
despite extensive investigations, are still not fully understood.
The Fourier power density spectrum (PDS) of GRBs, on the other
hand, seem to show relatively simple behavior. \cite{Giblin1998}
found that a typical PDS shows a low-frequency power-law component
and a high-frequency flat component (usually associated with
Poisson noise). \cite{Beloborodov1998, Beloborodov2000} considered
each GRB as a realization of some common stochastic process and
showed that when averaged over many bursts the resulting PDS
exhibits a power-law behavior with an exponent of -1.67, which the
authors note is consistent with the -5/3 Kolmogorov spectral index
expected from processes involving turbulent flow. In addition,
they claim that there is a break in the averaged PDS at $\sim$ 1
Hz. The authors were not in a position to correct their sample for
the time dilation due to the cosmological redshifts.

\cite{Lazzati2002} analyzed GRB power spectra by dividing them
into six luminosity bins using the variability-luminosity
correlation~\citep{Fenimore2000,Reichart2001,Guidorzi2005a,
Guidorzi2005b, Guidorzi2006, Li2006, Rizzuto2007}. The PDS was
averaged in each bin after correcting for pseudo-redshifts
obtained through the variability-luminosity
relation~\citep{Fenimore2000}. \cite{Lazzati2002} showed that the
dominant frequency ($f_d$) of the PDS is strongly correlated with
the variability parameter obtained by taking a modified variance
of the de-trended light curve~\citep{Fenimore2000}. Here $f_d$ is
obtained by finding the maximum of the function $f \times
\rm{PDS}$$(f)$ (see \cite{Lazzati2002} for more details). The
author further states that the red-noise component of the averaged
PDS for the six luminosity bins is well described by a broken
power-law function with a low-frequency slope of -2/3 and
high-frequency slope of -2. In this case, the break frequency is a
function of both luminosity and the variability parameter.

\cite{Borgonovo2007} did a similar analysis of power spectra but
used measured redshift information to correct for time dilation
effects before averaging. The burst sample was subdivided into two
populations based on the calculated values of the autocorrelation
function. After averaging, the PDS of one population shows a
power-law index of $\sim$ -2.0 (consistent with the spectral index
expected of Brownian motion) and the PDS of the other population
is characterized by a low-frequency exponentially decaying
component and a high-frequency power-law component with an index
of $\sim$ -1.6 (which again is consistent with the -5/3 Kolmogorov
spectral index).

Most of the previous work on power density spectra of GRBs has
been based on observations with the Burst and Transient Source
Experiment (BATSE) on the Compton Gamma Ray Observatory
\citep{Giblin1998,Beloborodov1998,Beloborodov2000,Lazzati2002,Borgonovo2007}
where a relatively modest amount of redshift information is
available. The launch of the $Swift$ satellite~\citep{Gehrels2004}
ushered in a new era of GRB research. Due to its rapidly
disseminated, arcsecond GRB positions, $Swift$ has enabled more
subsequent redshift measurements of GRBs than ever before. The
availability of redshift information enables the study of
rest-frame properties of bursts and provides an opportunity for
further exploration of correlations involving burst parameters
such as luminosity and variability.

In this paper we present a study of Fourier power density spectra
of 206 $Swift$ long bursts. Unlike previous work, we avoid
averaging PDS of multiple bursts and examine them individually. We
have developed a method to estimate the uncertainties in PDS for
each burst. Then we extract PDS for all the GRBs in the sample and
investigate the distribution of the extracted parameters. The
structure of the paper is the following: In
section~\ref{data_analysis} we discuss our methodology for
extracting the PDS. In section~\ref{results}, we present our
results for a sample of 206 long bursts and investigate various
correlations between the extracted parameters. In addition, we
propose a new frequency-luminosity relation based on a sample of
58 GRBs with spectroscopically measured redshifts. In
section~\ref{discussion}, we discuss observational biases of our
results. Finally, in section~\ref{conclusion} we summarize our
conclusions. In this work, we have adopted the standard values for
the cosmological parameters: $\Omega_M = 0.27$, $\Omega_L = 0.73$
and the Hubble constant $H_0$ is $70\,\rm(kms^{-1})/Mpc$.
Throughout this paper, the quoted uncertainties are at the 68\%
confidence level, unless noted otherwise.

\section[]{Data Analysis} \label{data_analysis}

\subsection[]{Light Curve Extraction} \label{light_curve}

$Swift$ BAT is a highly sensitive, coded aperture
instrument~\citep{barthelmy2005}. BAT uses the shadow pattern
resulting from the coded mask to facilitate few arc-minute
localization of gamma-ray sources. In order to generate background
subtracted light curves, we used a process called mask weighting.
The mask weighting assigns a ray-traced shadow value for each
individual event, which then enables the user to calculate light
curves or spectra.

We used the \texttt{batmaskwtevt} and \texttt{batbinevt} tasks in
the $Swift$ BAT FTOOLS to generate mask weighted,
background-subtracted light curves in the BAT energy range
$15-200$ keV. The light curves that are generated have rates that
are measured in counts per second per detector ($\rm
counts/sec/det$). In addition, the above tools also generate
uncertainties associated with the rates that are calculated by
propagation of errors from raw counts (subject to Poissonian
noise). For the BAT instrument one can potentially go down to the
minimum time binning of $\sim$ 0.1 ms. However, in this work, we
used 1-ms time binned light curves.

\subsection[]{Fourier Analysis} \label{FourierAnalysis}

We calculate the Fourier transform, $a_f$, of each GRB light
curve, $R(t)$ (measured in counts/sec/detector), using a standard
Fast Fourier Transform (FFT)\footnote{We used the FFT routine in
the IDL (Interactive Data Language) data analysis package.
\\
\texttt{http://www.ittvis.com/ProductServices/IDL.aspx}}
algorithm~\citep{JenkinsWatts1969, Press2002}. We used a time
segment of the burst light-curve where the total fluence is
accumulated (i.e. start and end times corresponding to burst T100
which is calculated by the \texttt{battblocks} task). The PDS of
each burst is calculated using $P_f=a_f \, a_f^*$. The power
spectra are not normalized nor are they averaged. In addition, we
have employed logarithmic binning for our power spectra.

This process of treating PDS individually is different from that
of \cite{Beloborodov1998,Beloborodov2000,Lazzati2002} and
\cite{Borgonovo2007}, as they used some averaging process to
obtain the slope of the red-noise component of the power spectra.
The wide variety of light curves exhibited by GRBs is potentially
indicative of different emission and scattering processes that
eventually shape the observed light curves and therefore we have
avoided averaging power spectra so as not to compromise this
valuable information.

\begin{figure}
\includegraphics[width=84mm]{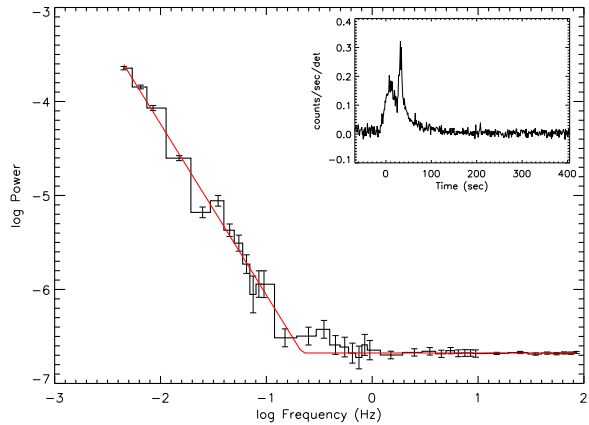}
\caption{Power Density Spectrum of GRB 081203A. The low frequency
power-law component is referred to as the `red-noise' component
and the flat high frequency region is called the `white-noise'
component. The inset shows the light curve of
GRB081203A.}\label{power_spectrum}
\end{figure}

The uncertainties of the individual PDS are calculated as follows:
For each burst, we simulate 100 light curves based on the original
light curve ($R^{{\rm real}}_{{\rm bin}}$) and its uncertainty
($R^{{\rm real \,error}}_{\rm bin}$), i.e.
\begin{equation}\label{eq:no1}
R^{{\rm simulated}}_{{\rm bin}} = R^{{\rm real}}_{{\rm bin}} +
\zeta \times R^{{\rm real \,error}}_{\rm bin}.
\end{equation}
Here $\zeta$ is a random number generated from a gaussian
distribution with the mean equal to zero and the standard
deviation equal to one. For each simulated light curve we
calculate a PDS. Then we re-bin each PDS logarithmically. The
uncertainties in the original PDS (obtained from the original
light curve) are derived by taking the standard deviation of the
100 simulated PDS.

The power spectra for the GRBs in the sample are fitted with the
function depicted in equation~\ref{eq:no2} (see
figure~\ref{power_spectrum} for a typical fit). This function
consists of a power-law component (to fit the low-frequency ``red
noise'' component) and a constant component (to fit the flat
high-frequency ``white noise'' component).
\begin{equation}\label{eq:no2}
\log P(f) = \left\{
\begin{array}{lc}
\alpha (\log f-\log f_{\rm th})+\log P_{\rm w} & {\rm for}\ f\le f_{\rm th}\\
\log P_{\rm w} & {\rm for}\ f>f_{\rm th}.
\end{array}
\right.
\end{equation}
Here $f_{\rm th}$ is the threshold frequency where the red-noise
component intersects the white-noise component of the power
density spectrum, and $P_{\rm w}$ is the white-noise power
density.

\section{Results} \label{results}

\begin{figure}
\includegraphics[width=84mm]{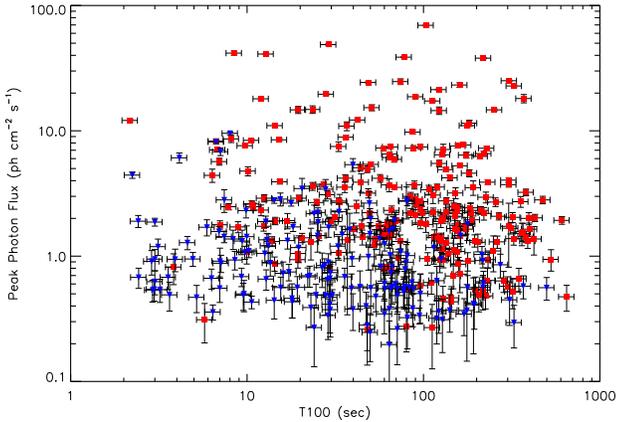}
\caption{The peak photon flux versus the T100 duration. The final
sample with a significant red noise component is shown in red
boxes and the sample with no or weak red-noise component is shown
in blue diamonds. Note that conservative 10\% uncertainties were
assumed for T100 values.}\label{GoodBadSample}
\end{figure}

Out of 451 GRBs which triggered $Swift$ BAT from 2004 December 19
to 2009 December 31, we selected a sample of 226 long GRBs that
show a significant red-noise component above the flat white-noise
region. In figure~\ref{GoodBadSample}, we represent the two
samples (the sample with clear red noise component is shown in red
boxes and the sample with no or weak red-noise component is shown
in blue inverted triangles) in a peak-photon-flux versus
T100-duration plot. For the most part, the bursts that do not show
a clear red-noise component are generally either weak and/or short
in duration.

\begin{figure}
\includegraphics[width=84mm]{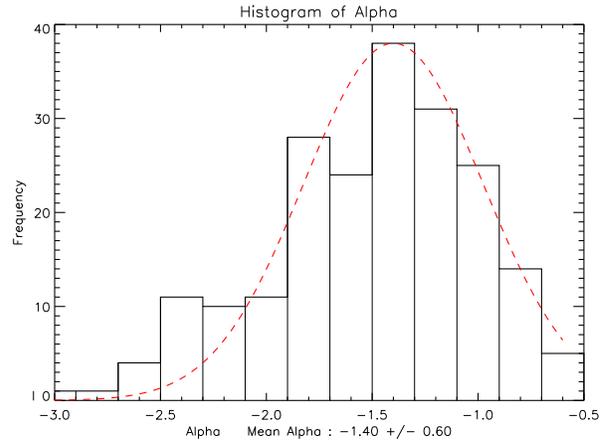}
\caption{Histogram of the extracted slopes ($\alpha$). The
distribution shows a peak around $-1.4 \pm
0.6$.}\label{fft_alpha_histogram}
\end{figure}

\begin{figure}
\includegraphics[width=84mm]{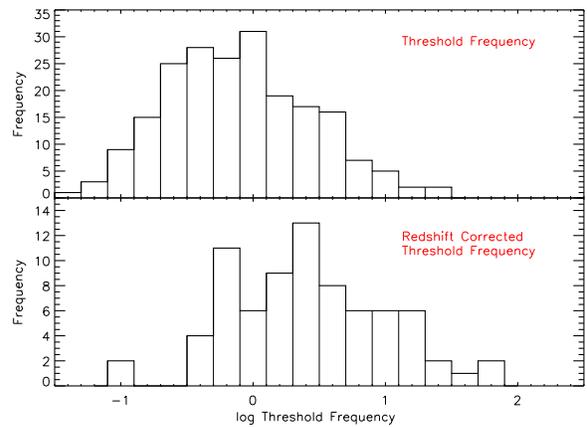}
\caption{A histogram of extracted threshold frequencies ($f_{\rm
th}$) is shown in the top panel. The histogram peaks around $\sim$
1 Hz. The bottom panel shows a histogram of redshift-corrected
$f_{\rm th}$ for the subset of bursts with redshift measurements.
Both distributions show a large
dispersion.}\label{fft_th_freq_histogram}
\end{figure}

\begin{figure}
\includegraphics[width=84mm]{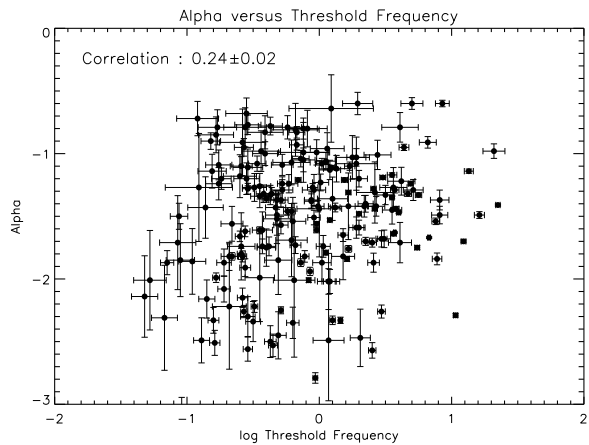}
\caption{The extracted slope, $\alpha$ as a function of the
threshold frequency, $f_{\rm th}$. A very weak positive
correlation is observed.}\label{fft_alpha_vs_freq}
\end{figure}

For the selected sample of 226 long GRBs we fitted the
corresponding PDS with a simple power-law behavior given in
equation~\ref{eq:no2}, using the nonlinear least squares routine
MPFIT~\citep{Markwardt2009}. A typical fit is shown in
figure~\ref{power_spectrum}. Out of the 226 GRBs in the sample, 20
bursts could not be fitted by a simple power-law. These GRBs were
excluded from further analysis. For the final sample of 206
bursts, the distributions of the extracted slopes ($\alpha$) and
threshold frequencies ($f_{\rm th}$) are shown in
figure~\ref{fft_alpha_histogram} and
figure~\ref{fft_th_freq_histogram} respectively. The distribution
of slopes ($\alpha$'s) has a Gaussian-like shape and peaks around
$\sim\,$-1.4 with $\sigma$ of about 0.6. The distribution of
threshold frequencies ($f_{\rm th}$'s) peaks around 1 Hz and also
shows a broad distribution. The distribution of the redshift
corrected $f_{\rm th}$ (i.e. $f_{\rm th} (1+z)$), as depicted in
the bottom panel of figure~\ref{fft_th_freq_histogram}, shows a
large dispersion and non-gaussian shape. In
figure~\ref{fft_alpha_vs_freq}, we show a plot of $\alpha$ and
$f_{\rm th}$; we see a very weak positive correlation ($0.24 \pm
0.02$) but we note at this stage of the analysis that $f_{\rm th}$
has not been corrected for noise contamination nor has it been
corrected for redshift.

\begin{figure}
\includegraphics[width=84mm]{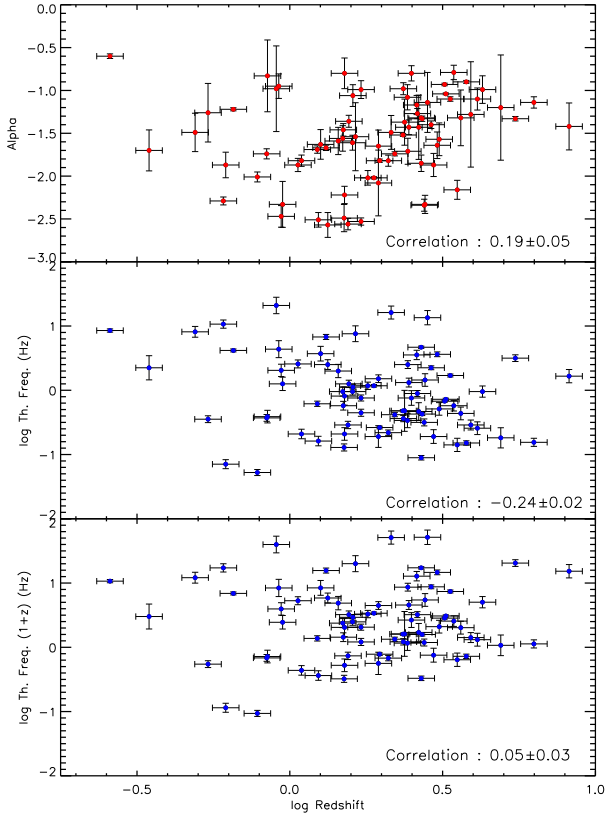}
\caption{The slope $\alpha$ as a function of redshift (top panel),
the threshold frequency as a function of redshift (middle panel),
and redshift-corrected $f_{\rm th}$ as a function of redshift
(bottom panel). The top and the middle panels show very weak
correlations while the bottom panel shows no significant
correlation. Note that conservative 10\% uncertainties were
assumed for redshift values. }\label{FreqVsRedshift}
\end{figure}

There are $76$ GRBs in our sample with measured redshifts
(spectroscopic or otherwise). For this sub-sample it is
interesting to see whether the extracted parameters are redshift
dependent. Figure~\ref{FreqVsRedshift} shows $\alpha$ (top panel),
$f_{\rm th}$ (middle panel), and the redshift-corrected $f_{\rm
th}$ (bottom panel) as a function of redshift. Very weak
correlations are observed between $\alpha$ and redshift and also
between $f_{\rm th}$ and redshift. However, no significant
correlation is observed between $f_{\rm th}(1+z)$ and redshift.

In \cite{Ukwatta2009var} we proposed a correlation between the
isotropic peak luminosity and the red-shift corrected $f_{\rm th}$
based on 27 GRBs. To investigate this further with a larger sample
we have selected a sample of 58 bursts with spectroscopically
measured redshifts and good spectral information. For this sample,
we have calculated isotropic peak luminosity as described
in~\cite{Ukwatta2010lag}. Based on the availability of spectral
information, we have divided the sample into three sub-samples:
``Gold'', ``Silver'', and ``Bronze''. The ``Gold'' sample with 15
bursts have all Band spectral parameters
measured~\citep{band1993}. In the ``Silver'' sample (15 bursts),
the $E_{\rm p}$ has been determined by fitting a cutoff
power-law\footnote{$dN/dE \sim E^\alpha \exp{(-(2+\alpha)E/E_{\rm
p})}$} (CPL) to spectra. These 15 bursts do not have the
high-energy spectral index, $\beta$, measured, so we used the mean
value of the BATSE $\beta$ distribution, which is $-2.36\pm0.31$
\citep{Kaneko2006, Sakamoto2009}. The ``Bronze'' sample, with 28
bursts, does not have a measured $E_{\rm p}$. We have estimated it
using the power-law index ($\Gamma$) of a simple power-law (PL)
fit as described in \cite{Sakamoto2009}. For these 28 bursts, the
low-energy spectral index, $\alpha$, and the high-energy spectral
index, $\beta$, were not known, so we used the mean values of the
BATSE $\alpha$ and $\beta$ distribution, which are $-0.87\pm0.33$
and $-2.36\pm0.31$ respectively \citep{Kaneko2006, Sakamoto2009}.

\begin{figure}
\includegraphics[width=84mm]{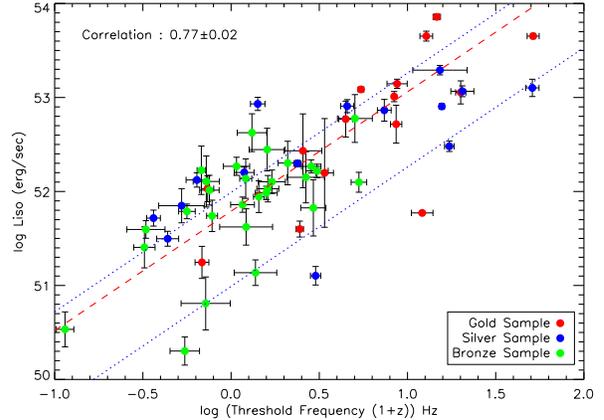}
\caption{Isotropic peak luminosity as a function of the redshift-
corrected threshold frequency, $f_{\rm th}(1+z)$. The parameters
are correlated with a correlation coefficient of $0.77 \pm 0.02$
and the best-fit power-law yields an exponent of $1.27 \pm
0.12$.}\label{fth_vs_Liso}
\end{figure}

The isotropic luminosity as a function of the redshift-corrected
threshold frequency is shown in figure~\ref{fth_vs_Liso}. In the
figure, the ``Gold'', ``Silver", and ``Bronze" samples are shown
in red, blue, and green filled circles respectively. A clear
positive correlation can be seen in the figure. The Pearson's
correlation coefficient is $0.77 \pm 0.02$, where the uncertainty
was obtained through a Monte Carlo simulation. The probability
that the above correlation occurs due to random chance is $\sim \,
4.5 \times 10^{-9}$. Our best-fit is shown as a red dashed line in
figure~\ref{fth_vs_Liso} yielding the following relation between
$L_{\rm iso}$ and $f_{\rm th}$:
\begin{equation}\label{eq:no3}
\log L_{\rm iso} = (51.79 \pm 0.12) + (1.27 \pm 0.12) \log (f_{\rm
th} (z+1)).
\end{equation}
To compensate for the large scatter in the plot, the uncertainties
of the fit parameters are multiplied by a factor of $\sqrt{\rm
\chi^2/ndf} = \sqrt{1766/56} \approx 6.0$ where $\rm ndf$ is the
number of degrees of freedom. The blue dotted lines indicate the
estimated 1$\sigma$ confidence level, which is obtained from the
cumulative fraction of the residual distribution taken from 16\%
to 84\%.

Our result for the slope in figure \ref{fth_vs_Liso} is consistent
with the value of $1.4 \pm 0.2$ obtained by \cite{Ukwatta2009var}
using 27 GRBs. This is encouraging because the results of
\cite{Ukwatta2009var} were obtained using non-mask-weighted
event-by-event data instead of the mask-weighted data that we use
in the current work. We also note that with the increase of the
sample size by about a factor of two the correlation coefficient
has increased from $0.69 \pm 0.03$ to $0.77 \pm 0.02$. The
correlation between frequency and luminosity is clearly intriguing
but there remain observational biases which we address in a later
section.

\begin{figure}
\includegraphics[width=84mm]{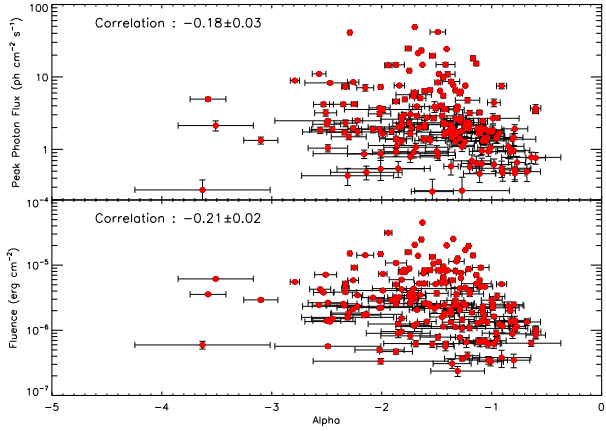}
\caption{The peak photon flux as a function of alpha (top panel)
and the fluence as a function of alpha (bottom panel). Very weak
negative correlations are observed in both cases.
}\label{SlopeVsBrightness}
\end{figure}

It has been reported previously~\citep{Beloborodov2000} that the
PDS slope is correlated with the burst brightness. In order to
check our sample for this effect, we display, in
figure~\ref{SlopeVsBrightness}, the slope ($\alpha$) against
brightness indicators: the peak photon flux and the fluence. Very
weak negative correlations are observed in both cases.

\begin{figure}
\includegraphics[width=84mm]{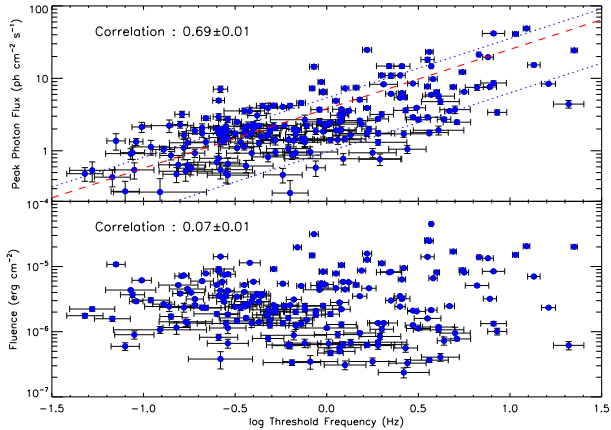}
\caption{The peak photon flux as a function of threshold frequency
(top panel) and the photon fluence as a function of threshold
frequency (bottom panel). No significant correlation is observed
between fluence and $f_{\rm th}$ but a significant correlation is
observed between peak photon flux and $f_{\rm th}$.
}\label{FreqVsBrightness}
\end{figure}

The other extracted parameter, the noise crossing threshold
frequency ($f_{\rm th}$) of the PDS, is also expected to depend on
the brightness of the GRB. Presumably, the `red-noise' component
of the PDS comes primarily from the GRB but the flat `white-noise'
component can in principle arise from the Poisson noise (intensity
fluctuations) associated with the GRB and the natural background
in the field-of-view of the detector. For distant and/or
intrinsically weak bursts, noise unrelated to the burst may
dominate the observed white-noise component, thereby overwhelming
the red-noise part of the signal. This, in turn, would make the
extraction of the threshold frequency brightness-dependent. In
figure~\ref{FreqVsBrightness} we plot the peak photon flux and the
photon fluence as a function of the threshold frequency. The red
dashed line in the top panel of figure~\ref{FreqVsBrightness} is
the best fit, given by equation~\ref{eq:no3.5}, and blue dotted
lines indicate a $1 \sigma$ confidence interval.
\begin{equation}\label{eq:no3.5}
\log {\rm PPF} = (0.58 \pm 0.03) + (0.82 \pm 0.04) \log f_{\rm th}
\end{equation}
Indeed, a positive correlation can be seen between $f_{\rm th}$
and the peak photon flux. However, no significant correlation is
observed between $f_{\rm th}$ and fluence. We discus this
important matter further in the next section.

\section{Discussion} \label{discussion}

\begin{figure}
\includegraphics[width=84mm]{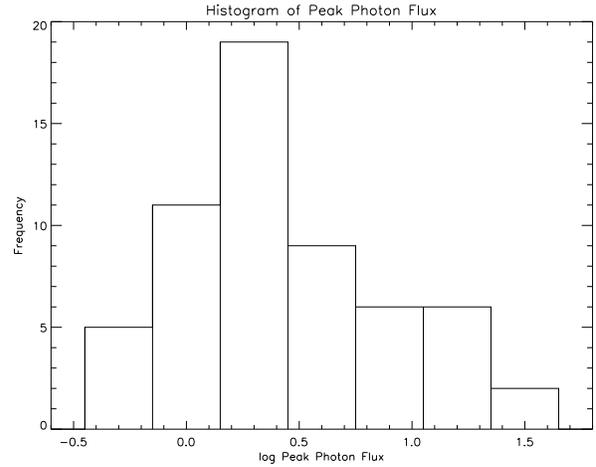}
\caption{Histogram of the peak photon flux of the sample of 58
bursts used to generate
figure~\ref{fth_vs_Liso}.}\label{ppf_histogram}
\end{figure}

\begin{figure}
\includegraphics[width=84mm]{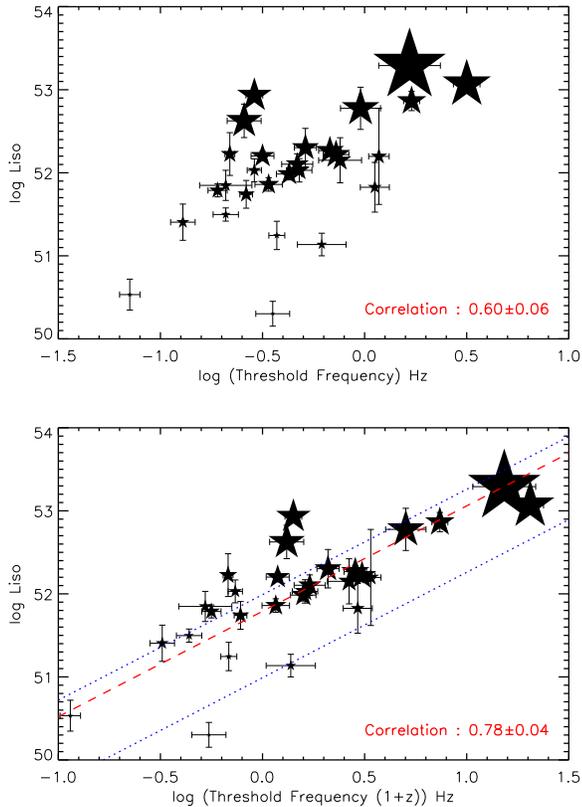}
\caption{The isotropic peak luminosity as a function of threshold
frequency (with and without redshift correction) for a sample of
bursts with narrow apparent brightness range. The red-dashed and
blue-dotted lines are the same fit curves shown in
figure~\ref{fth_vs_Liso}. The size of the star is proportional to
the redshift of the burst.}\label{fth_ppf_freq}
\end{figure}

It is conceivable that the proposed frequency-luminosity
correlation is a direct result of the observed correlation between
$f_{\rm th}$ and the peak photon flux of the burst (see the top
panel of figure~\ref{FreqVsBrightness}). If this is the case, then
for a statistically significant sample of bursts with similar
apparent brightness, we should not see a correlation between
$f_{\rm th}$ and $L_{\rm iso}$. In order to select a sample of
GRBs with similar apparent brightness we plot in
figure~\ref{ppf_histogram} the peak photon flux distribution for
the sample of 58 bursts used to investigate the
frequency-luminosity correlation in figure~\ref{fth_vs_Liso}. We
see from figure~\ref{ppf_histogram} that about half of the sample
(28 GRBs) have a very similar peak photon flux (0.0 $< \, \log$
(peak photon flux) $<$ 0.5). For this subset of bursts we plotted
their peak photon flux and $L_{\rm iso}$ as a function of $f_{\rm
th}$ and the results are shown in figure~\ref{fth_ppf_freq}. In
the top panel of figure~\ref{fth_ppf_freq}, it is clear that there
is a significant correlation between $f_{\rm th}$ and $L_{\rm
iso}$ with correlation coefficient of $0.60 \pm 0.06$. This
implies that the correlation observed in figure~\ref{fth_vs_Liso}
($f_{\rm th}(1+z)$ - $L_{\rm iso}$ correlation) is not entirely
due to the correlation seen in the top panel of
figure~\ref{FreqVsBrightness} ($f_{\rm th}-$peak photon flux
correlation). We now correct the $f_{\rm th}$ of this limited
sample (with similar apparent brightness) for redshift to see its
effect. Plotted in the bottom panel of figure~\ref{fth_ppf_freq}
are the redshift corrected data. We note that the correlation
strength increases to a value of $0.78 \pm 0.04$, in part due to
the natural correlation between redshift and $L_{\rm iso}$.

\begin{figure}
\includegraphics[width=84mm]{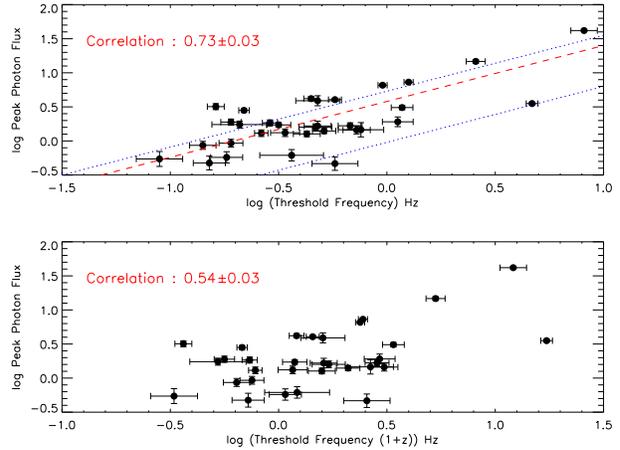}
\caption{Peak photon flux as a function of $f_{\rm th}$ (top
panel) and redshift corrected $f_{\rm th}$ (bottom panel) for a
sample bursts with roughly constant luminosity. The red dashed
line in the top panel is the best fit line obtained in
figure~\ref{FreqVsBrightness} and blue dotted lines indicate $1
\sigma$ confidence interval.}\label{fth_ppf_const_Liso}
\end{figure}

In addition, we can approach the issue from the other direction,
i.e., we select a subset of bursts with similar luminosity and ask
the question whether the correlation between peak photon flux and
$f_{\rm th}$ comes from the proposed $f_{\rm th}(1+z)$ - $L_{\rm
iso}$ correlation. In order to perform this test we selected a
subset of bursts which have the roughly the same $L_{\rm iso}$
values (51.5 $< \, \log L_{\rm iso} < $ 52.5) and plotted their
peak photon flux as a function of $f_{\rm th}$. In
figure~\ref{fth_ppf_const_Liso}, we show the peak photon flux as a
function of $f_{\rm th}$ (top panel) and the redshift corrected
$f_{\rm th}$ (bottom panel). There is clearly a strong correlation
between the two parameters in both panels. It is interesting,
however, that after the redshift correction, the correlation
strength drops significantly. Accordingly, it would appear that
the correlation between the peak photon flux and $f_{\rm th}$ (top
panel of figure~\ref{FreqVsBrightness}) is not entirely due to the
$f_{\rm th}(1+z)$ - $L_{\rm iso}$ correlation
(figure~\ref{fth_vs_Liso}). Since the spectral power is
proportional to the square of the flux and the PDS follows a
$f^{-\alpha}$ behavior (see figure~\ref{fft_alpha_histogram}), we
expect to see a correlation between peak photon flux and $f_{\rm
th}$. Hence, this correlation is in most part observational.

Now we turn to the question of the dependance of the extracted
threshold frequency on the noise-level of the burst. The obvious
question is how to determine the noise-level for each burst. One
way of defining the noise-level is the following:

\begin{equation}\label{eq:no4}
\rm Noise \, Level = \frac{Std. Dev. (Detrended \, LC)}{Peak
\,Count \,Rate} \times 100 \%.
\end{equation}

The detrending of the light curve (LC) can be done in a number of
ways and we adopted the following method. We generated two light
curves of the same burst with two bin sizes. In order to produce
the coarser binned light curve, we chose a time bin size that
resulted in at least 100 points in the burst duration (T100). The
other light curve may have bin sizes that vary from 1 ms up to the
coarser bin size. Clearly, with the different binning, the two
light curves will have a different number of points. In order to
properly detrend, we need to have the same number of points in the
two light curves. We accomplish this by using a simple linear
interpolation of the coarser binned light curve. The interpolated
light curve is then subtracted  from the finer binned light curve
to generate the detrended light curve.

Using equation~\ref{eq:no4} we extract a noise level for each
burst. However, the extracted noise-level depends on the bin size
used in the detrending process. This aspect needs to be either
removed or accounted for before the noise level of all the bursts
can be treated on an equal footing.

The level of the flat white-noise region of the PDS does not
depend on the bin size, i.e., for a given burst the white noise
level is constant irrespective of the bin size, and for that
matter so too are the extracted parameters $\alpha$ and $f_{\rm
th}$. In order to remove the bin size dependence in the extraction
of the noise-level, we modify equation~\ref{eq:no4} as follows.
\begin{equation}\label{eq:no5} \rm Noise \, Level =
\frac{1}{\sqrt{N}}\frac{Std. Dev. (Detrended \, LC)}{Peak \,Count
\,Rate} \times 100 \%.
\end{equation}
Here $N$ is the number of data points in the finely binned light
curve. Our tests indicate that the results given by
equation~\ref{eq:no5} do not depend on the time bin size of the
light curve and provide a robust measure of the noise-level of a
given burst.

\begin{figure}
\includegraphics[width=84mm]{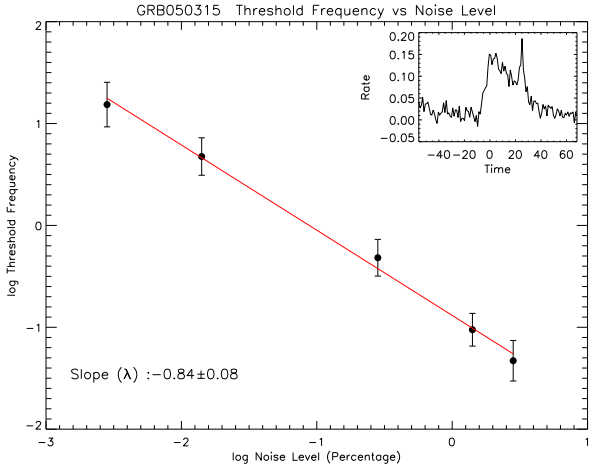}
\caption{The extracted threshold frequency as a function of the
noise-level in a log-log scale. The threshold frequency displays a
power-law dependance on the noise-level of the burst with a index
($\lambda$) of $-0.84 \pm 0.08$ for GRB 050315. The inset shows
the time profile of the burst.}\label{Simulated_LCs}
\end{figure}

In order to further investigate the dependance of $f_{\rm th}$ on
the noise-level, we performed additional tests. We simulated
different noise levels by adding increasing amounts of Gaussian
noise to a burst light curve (in this case GRB 050315). Then we
extracted $f_{\rm th}$ values for each setting of the noise-level.
Our results, the extracted frequency values versus the
noise-level, are shown in figure~\ref{Simulated_LCs} as a log-log
plot. The threshold frequency does indeed depend on the
noise-level.  However, there is a linear relationship between the
logarithmic values of the two quantities. This relation is
important to know because it can be used to correct the extracted
$f_{\rm th}$ values to some nominal noise-level that is common to
all bursts in the sample.

\begin{figure}
\includegraphics[width=84mm]{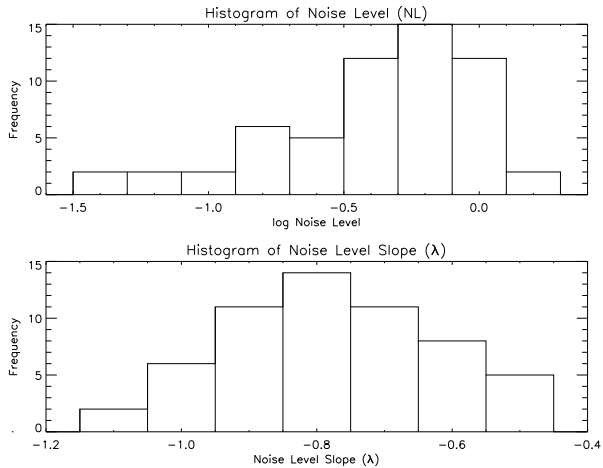}
\caption{Distribution of noise-levels (top panel) and noise-level
slope, $\lambda$, (bottom panel) in the sample.
}\label{nl_and_alpha_histogram}
\end{figure}

By performing the same test on the other bursts in our sample, we
established that the relation between $f_{\rm th}$ and the
noise-level depends on the profile of the burst, i.e., the slope
($\lambda$) of the log-log plot is different for each burst. Shown
in the bottom panel of figure~\ref{nl_and_alpha_histogram} is the
distribution the slopes, $\lambda$, obtained for our sample of 58
bursts used in the $f_{\rm th}$-$L_{\rm iso}$ relation.
Correspondingly, the noise-level ($\rm NL$) distribution is shown
in the top panel of figure~\ref{nl_and_alpha_histogram}. This
distribution shows a clear peak around the log value of -0.2 ($\rm
NL \sim 0.6$) while the $\lambda$ distribution peaks around the
value of -0.8.

\begin{figure}
\includegraphics[width=84mm]{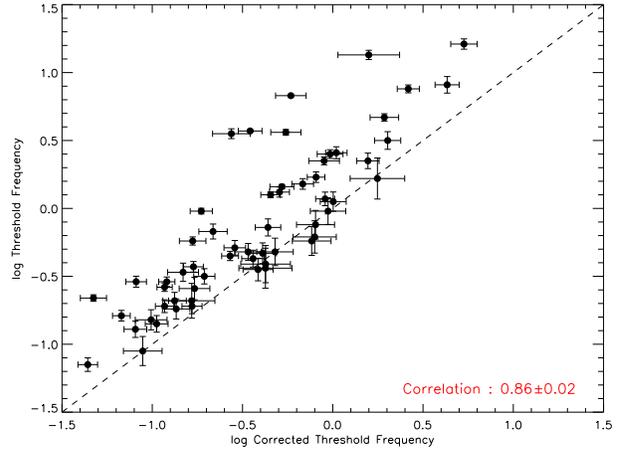}
\caption{The threshold frequency versus the noise-corrected
threshold frequency. The two parameters show a strong correlation
with correlation coefficient of 0.86. The dashed line indicates
the equality line of the two
parameters.}\label{th_freq_vs_corrected_th_freq}
\end{figure}

\begin{figure}
\includegraphics[width=84mm]{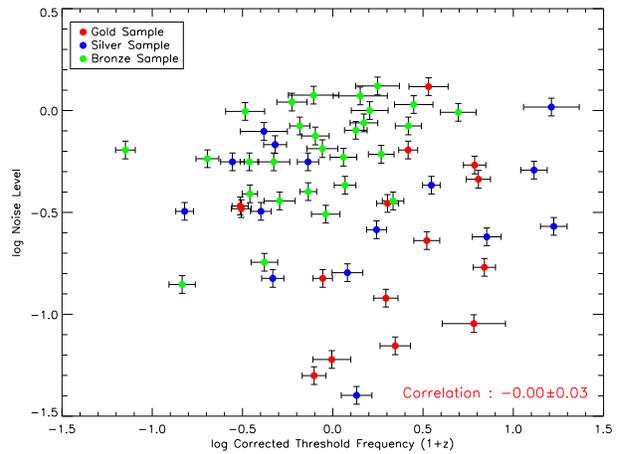}
\caption{The noise-corrected, redshift-corrected threshold
frequency versus noise
level.}\label{Corrected_f_th_versus_noise_level}
\end{figure}

We are now in a position to treat all the bursts in our sample on
an equal footing and test whether the $f_{\rm th}$-$L_{\rm iso}$
correlation, observed in figure~\ref{fth_vs_Liso}, survives. The
aim is to extract threshold frequencies which are consistent with
a noise-level that is common to all the bursts in our sample. In
order to accomplish this, we choose an arbitrary noise level of
$\rm NL=1.0$ (see figure~\ref{Simulated_LCs}) and use the
following relation to extract a corrected $f_{\rm th}$ for each
burst:
\begin{equation}\label{eq:no6}
\log f_{\rm th [NL=1]} = \log f_{\rm th [NL=burst]} - \lambda_{\rm
burst} \log (\rm NL_{[\rm NL=burst]}).
\end{equation}
Here, $f_{\rm th [NL=burst]}$ is the extracted threshold frequency
for a given burst, $\lambda_{\rm burst}$ is the noise-level slope
corresponding to the same burst and $\rm NL_{[NL=burst]}$ is the
burst noise-level determined by equation~\ref{eq:no5}. The
correction procedure is repeated for each burst in our sample. To
gauge the size of the correction, we plot in
figure~\ref{th_freq_vs_corrected_th_freq} (in a log-log scale) the
corrected $f_{\rm th}$ values versus the uncorrected $f_{\rm th}$.
We note that there is a strong correlation between the two
parameters. This is a reflection of the clustering of the $\rm NL$
and $\lambda$ seen in figure~\ref{nl_and_alpha_histogram}. We also
plotted the NL as a function of the noise-corrected,
redshift-corrected $f_{\rm th}$ in
figure~\ref{Corrected_f_th_versus_noise_level}. There is no
correlation between these two parameters, thus giving us
confidence in the noise correction procedure.

\begin{figure}
\includegraphics[width=84mm]{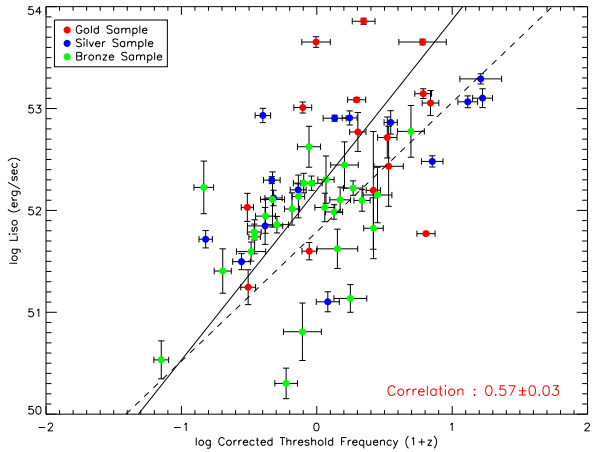}
\caption{The noise-corrected, redshift-corrected threshold
frequency versus isotropic peak luminosity. The correlation
coefficient between the two parameters is $0.57 \pm 0.03$. The
solid line shows the best-fit power-law with a index of $1.67 \pm
0.01$. The dashed line shows the best-fit from
figure~\ref{fth_vs_Liso}.}\label{f_th_versus_Liso}
\end{figure}

We show in figure~\ref{f_th_versus_Liso} the noise-corrected
threshold frequency - luminosity relation. As is evident, the
relation survives the noise correction albeit with a somewhat
smaller correlation coefficient of $0.57 \pm 0.03$. Various
correlation coefficients of the relation are shown in
Table~\ref{table}, where the uncertainties were obtained through a
Monte Carlo simulation. The null probability that the correlation
occurs due to random chance is also given for each coefficient
type.

\begin{table}
\centering
 \begin{minipage}{84mm}
  \caption{Correlation coefficients \label{table}}
  \begin{tabular}{@{}lll@{}}
  \hline
  Coefficient Type     &  Correlation Coefficient & Null Probability \\
 \hline
Pearson's $r$          & 0.57$\pm$0.03 & $1.42\,\times\,10^{-5}$\\
Spearman's $r_{\rm s}$ & 0.58$\pm$0.04 & $1.72\,\times\,10^{-6}$\\
Kendall's $\tau$       & 0.43$\pm$0.03 & $2.03\,\times\,10^{-6}$\\
\hline
\end{tabular}
\end{minipage}
\end{table}

The new best-fit is shown as a solid line in
figure~\ref{f_th_versus_Liso} yielding the following relation
between $L_{\rm iso}$ and $f_{\rm th [NL=1]}$:

\begin{equation}\label{eq:no6}
\log L_{\rm iso} = (52.2 \pm 0.1) + (1.67 \pm 0.01) \log (f_{\rm
th} (z+1)).
\end{equation}
The uncertainties in the fitted parameters are expressed with the
factor of $\sqrt{\rm \chi^2/ndf} = \sqrt{1255/56} \approx 5.0$.

\section{Conclusion} \label{conclusion}

In this paper we have analyzed PDS of 206 GRBs. We fitted each PDS
with a simple power-law and determined the red-noise exponent and
the threshold frequency where white noise begins. For a subset of
GRBs, we extracted a frequency-luminosity relationship. For this
sample, we treated all bursts on an equal footing by determining a
common noise level, thereby minimizing the potential observational
biases. We summarize the main results of our analysis as follows:

\begin{itemize}
    \item The distribution of the extracted $\alpha$ (slope of
    the red-noise component) values peaks
    around -1.4 and that of $f_{\rm th}$ around 1 Hz.

    \item The dispersion in the distribution of $\alpha$
    is large and so the Kolmogorov index of -5/3 is accommodated by our
    analysis.

    \item The distribution of the redshift-corrected threshold frequency
    shows a large dispersion and is non-gaussian in
    shape.

    \item Evidence is presented for a possible frequency-luminosity
    relationship, i.e., the redshift-corrected $f_{\rm th}$ is correlated
    with the isotropic luminosity. The correlation coefficient is
    $0.57 \pm 0.03$ and the best-fit power-law has an index of $1.67
    \pm 0.01$. We appreciate that in reality there may be complicated underlying
    interrelationships involving peak photon flux, $f_{\rm th}$, and
    redshift and therefore the evidence for the frequency-luminosity
    relation should be considered tentative.

    \item The proposed frequency-luminosity correlation, if
    confirmed, may serve to provide a measure of the intrinsic variability
    observed in GRBs.

\end{itemize}

\section*{Acknowledgments}

We thank the anonymous referee for comments and suggestions that
significantly improved the paper. We also thank T. Sakamoto and C.
Guidorzi for useful discussions. The NASA grant NNX08AR44A
provided partial support for this work and is gratefully
acknowledged.

\end{document}